\documentclass{emulateapj}

\newcommand{\actaa}{Acta Astron.}

\shorttitle{Eclipsing binaries in the MACHO database}
\shortauthors{Derekas et al.}

\begin{document}

\title{Eclipsing binaries in the MACHO database: New periods and
classifications for 3031 systems in the Large Magellanic Cloud}

\author{A. Derekas\altaffilmark{1}, L. L. Kiss and T. R. Bedding}
\affil{School of Physics, University of Sydney, NSW 2006, Australia}

\altaffiltext{1}{E-mail: derekas@physics.usyd.edu.au}

\begin{abstract}

Eclipsing binaries offer a unique opportunity to determine fundamental physical
parameters of stars using the constraints on the geometry of the systems. Here
we present a reanalysis of publicly available two-color observations of about
6800 stars in the Large Magellanic Cloud, obtained by the MACHO project between
1992 and 2000 and classified as eclipsing variable stars. Of these, less than
half are genuine eclipsing binaries. We determined new periods and classified
the stars, 3031 in total, using the Fourier parameters of the phased light
curves. The period distribution is clearly bimodal, reflecting refer to the
separate groups of more massive blue main sequence objects and low mass red
giants. The latter resemble contact binaries and obey a period-luminosity
relation. Using evolutionary models, we identified foreground stars. The
presented database has been cleaned of artifacts and misclassified variables,
thus allowing searches for apsidal motion, tertiary components, pulsating stars
in binary systems and secular variations with time-scales of several years.

\end{abstract}


\section{Introduction}

The last decade witnessed the birth of a new research field, the large-scale
study of variable stars in external galaxies. This has been made possible by
the huge databases of microlensing observations in the Magellanic Clouds, such
as the MACHO \citep{alc93}, OGLE \citep{uda97} and EROS \citep{aub95} projects.
These programs (beyond their primary purpose) resulted in the discovery of
thousands of eclipsing binaries and many other types of variable stars and gave
an unprecedented homogenous coverage of their light curves. Recent all-sky
surveys also give a good opportunity to study large numbers of stars
\citep{pac06}. However, processing the huge amount of data can be quite
challenging, even when looking at seemingly simple issues such as
classification of variables, period determination, etc.

The astrophysical potential of large databases of eclipsing binaries has been
explored by a number of authors, but we are still far from a full exploitation.
Catalogs of eclipsing variables in the Magellanic Clouds have been published
with an increasing completeness \citep{alc97,wyr03,wyr04}. Attempts have been
made to improve our understanding of the formation of contact binaries,
especially those with giant components \citep{ruc97a,ruc97b,ruc98,mac99,ruc01}.
To help analyse large samples of stars, various automated pipelines have been
developed \citep{dev05,tam06,maz06}, while the large-number statistics helped
investigate orbital circularization over a broad range of stellar parameters
\citep{fac05}. Another application is determining accurate distances using
early-type detached eclipsing binaries, which has been shown to offer the most
accurate calibration of the local distance scale
\citep[e.g.,][]{gui98,rib00,wyi01,sal02,fit02,rib02,mic05}.

In contrast to these applications, the near-decade long time coverage of the
available data has rarely been utilised. For example, \citet{pal03} and
\citet{joh04} reported on searches for tertiary companions of eclipsing
binaries in the MACHO database, but no definite results have been published.

Here we present an analysis of the publicly available MACHO light curves of LMC
variables classified as eclipsing binaries, which have been online since 2001.
The main aim of this project is to measure period changes and search for
eclipsing binaries with pulsating components. The period-luminosity (P--L)
relations of contact binaries and their relation to pulsating red giants were
already discussed in \citet{der06}. In this paper we discuss the general
properties of the sample, presenting a full reclassification, newly determined
periods, the color-magnitude diagram and period-luminosity distributions.

\section{Period determination and classification}

The MACHO observations were carried out between 1992 and 2000 with the 1.27m
Great Melbourne telescope at Mount Stromlo Observatory, Australia. The
telescope was equipped with a specificly designed camera, which gave a 0.5
square degree field of view. The observations were obtained in two non-standard
bandpasses simultaneously: a 440--590 nm MACHO ``blue'' filter and a 590-780 nm
MACHO ``red'' filter \citep{coo95}. The typical number of observational
points obtained is about 1200. Some of the data are publicly available on the
MACHO website\footnote { The  MACHO variable star catalog is available at
http://wwwmacho.mcmaster.ca}, where one can select samples based on an
automatic variability type classification. Using the web-interface, we
individually downloaded light curves of all stars classified as eclipsing
binaries, 6833 in total. However, it became obvious very quickly that the
automatic classification was not perfect, as a large fraction of `eclipsing
binaries' turned out to be Cepheids, RR Lyrae stars or long-period variables.
We therefore needed to re-classify all stars. Furthermore, besides the problems
with the classification, we also found that the MACHO periods were incorrect
for significant number of stars. For this reason, we redetermined periods for
all the 6833 objects and classified them based on the light curve shapes. Since
the MACHO light curves contain more blue points than red ones, we used the
former for determining periods.

For an initial period determination, we used the Phase Dispersion Minimization
method \citep[PDM,][]{ste78}. The method is based on a technique to
minimize the sample variance of the phase diagram. It is calculated in the
following way. The phase diagram for a given trial period is divided into $m$
segments, each containing ${\rm n_{j}}$ ${\rm (j=1...m)}$ points. The variance
for each segment is defined as ${\rm s_{j}^{2}=\sum_{i=1}^{n_{j}} \left( m_{i}
- \overline{m}_{j} \right)^{2}/ \left( n_{j}-1 \right)}$, where ${\rm m_{i}}$
is the observed magnitude, ${\rm \overline{m}_{j}}$ is the mean magnitude:
${\rm \overline{m}_{j}=\sum_{i=1}^{n_{j}} m_{i}/n_{j}}$. The sample variance
${\rm s^{2}=\sum_{j=1}^{m} s_{j}^{2}}$ is minimized to get the best period. We
calculated 600 000 trial phase diagrams for each star, covering a wide range of
periods between 0.085 days and 1000 days (from 0.001 cycles/day to 12
cycles/day, with equidistant steps in frequency).

However, this was not enough for determining the true periods. As commonly
happens in finding periods from light curves of eclipsing binaries, a 
significant fraction of the PDM results were harmonics or subharmonics of the
true period (${\rm P_{\rm PDM}/P_{\rm true}}$ was a ratio of small integers).
To correct for this, we proceeded as follows. A visual inspection of every
phase diagram showed whether the actual period was an alias or just slightly
inaccurate.  In the case of an alias, we multiplied the PDM period by different
constants (in most cases by 2) until the shape of the curve was consistent with
that of an eclipsing binary. 

\begin{figure}
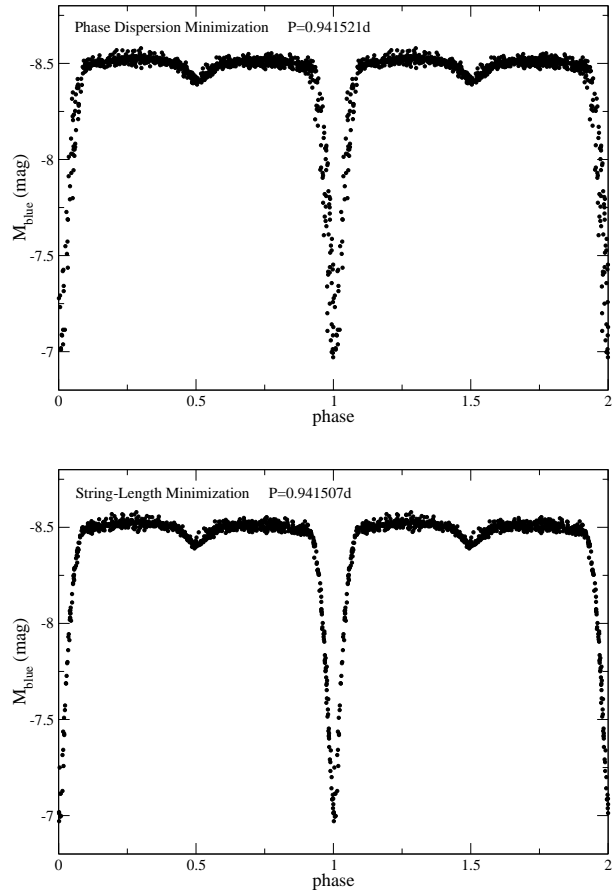

\includegraphics[width=8cm]{pdm_example.eps}   
\vskip5mm
\includegraphics[width=8cm]{sl_example.eps}
\caption[]{\label{pdm-sl} An example (81.9006.28) for improvement in period
determination using PDM (top panel) and SL (bottom panel) methods.}
\end{figure}

We next used the String-Length method  \citep{laf65, cla02} to improve period
determination. With this method, one calculates the total length of the phase
diagram for any given period as the following sum: ${\rm
SL^{2}=\sum_{i=1}^{n-1} \left( \left( m_{i+1}-m_{i}\right)^{2}+\left(
\varphi_{i+1}-\varphi_{i}\right)^{2}\right)}$, where ${\rm \left\{
\varphi_{i},m_{i} \right\}_{i=1}^{n}}$ is a folded dataset sorted in phase;
${\rm \varphi_{i}}$ and ${\rm m_{i}}$ are the phase and the magnitude of the
observations taken at time ${\rm t_{i}}$ (so that ${\rm
\varphi_{i}=\left[\left(t_{i}-t_{0}\right)/P\right]}$, where ${\rm t_{0}}$ was
chosen as an epoch of the deeper minimum, P is the period, [~] is the
fractional part). The best period here was the one that minimized ${\rm
SL^{2}}$. In our case, we applied the SL method for 1000 periods within
$\pm$1\% of the best PDM period. An example of the period improvement is shown
in Fig.\ \ref{pdm-sl}. The typical period improvements resulted in a change in
the 5-6th decimal place.

During the visual inspection of the phase diagrams we also made a rough
classification of all 6833 variables. Based on the light curve shape, phased
with the finally adopted periods, we placed each star into one of the following
categories: Algol-type, $\beta$ Lyrae-type, W UMa-type eclipsing binary,
pulsating star (including RR Lyraes, Cepheids, Miras, etc.), non-periodic or
multiply periodic, and unidentifiable. In cases of a pure sine-wave, which can
be observed in a contact binary or a pulsating star, we compared the amplitudes
of the MACHO blue and red phase diagrams. If there were noticeable color
variations exceeding a few hundredth of a magnitude, we considered the star as
pulsating variable.

After the whole procedure, we ended up with 3031 genuine eclipsing or
ellipsoidal binaries, the rest belonging to other type of variables. The
identification numbers, J2000 coordinates, calibrated colors and magnitudes,
periods and epochs of minimum light for the eclipsing variables are listed in
Table\ \ref{data} (available in full electronically). In the last column
of the table, we note cases where a very faint secondary eclipse might possibly
exist, which would result in a 1/2 ambiguity in the period. 12 of the 3031
systems have no periods because they showed signs of only one eclipse over the
eight years of MACHO observations.

\begin{figure}  
\epsscale{1.15}
\plotone{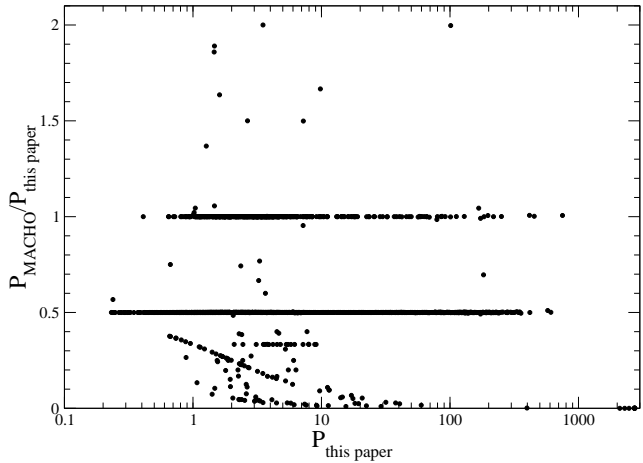}   
\caption[]{\label{comp} A comparison of periods presented in this study and
those of the MACHO project.} 
\end{figure}

A comparison of our periods with those in the online MACHO catalog shows that
there is a major disagreement. A plot of the ratios of the periods (Fig.\
\ref{comp}) reveals an interesting structure. A minor fraction of the period
ratios are scattered around 1, while the majority are dominated by the cases
when the MACHO period is half of ours. Also, we identified a number of stars
with period ratios at 1/3, 1/4, 1/5, etc. The diagonal line in the lower part
of the histogram corresponds to cases when the MACHO frequency (${\rm
1/P_{MACHO}}$) is ${\rm \pm 1}$ c/d offset from the true frequency or one of
its integer harmonics.

\begin{figure}  
\epsscale{1.15}
\plotone{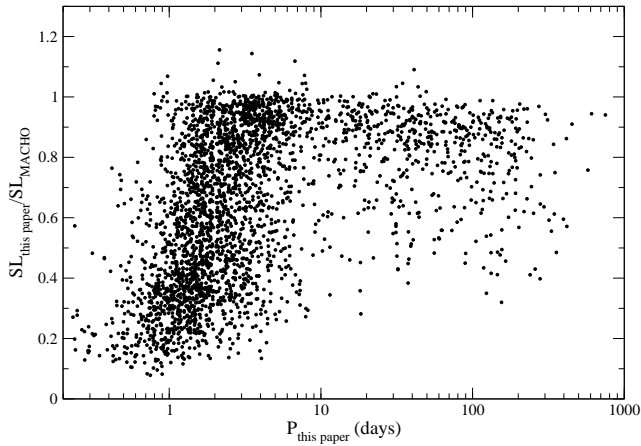}
\caption[]{\label{slratio} The ratio of the length of phase diagrams calculated
with period in this paper to that using the MACHO period versus period of this
paper. } 
\end{figure}

To illustrate the improvement of periods in this paper compared to those
in the online MACHO catalog, we performed the following statistical evaluation.
For each star, we calculated the string-length using both our period and that
of the MACHO online catalog. The ratio of the two lengths is plotted as a
function of period in Fig.\ \ref{slratio}. We see a considerable improvement
for all but a few stars; the points over 1.0 all refer to datasets with very
few points, for which none of the statistics is well defined.

Ratios other than 0.5 occurred almost exclusively in stars with eccentric
orbits, which causes a shift in the secondary minimum from 0.5 phase. These
problems clearly illustrate that one has to be very careful when determining
periods automatically for eclipsing binaries.

\begin{figure}  
\epsscale{1.15}
\plotone{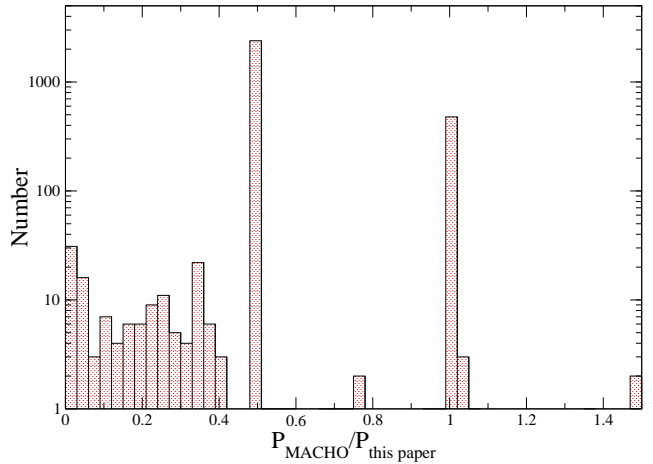}   
\caption[]{\label{hiscom} The histogram of period ratios presented in this study
and by the MACHO project. Note the vertical axis is in logarithmic scale.} 
\end{figure}

\begin{figure}
\epsscale{1.15}
\plotone{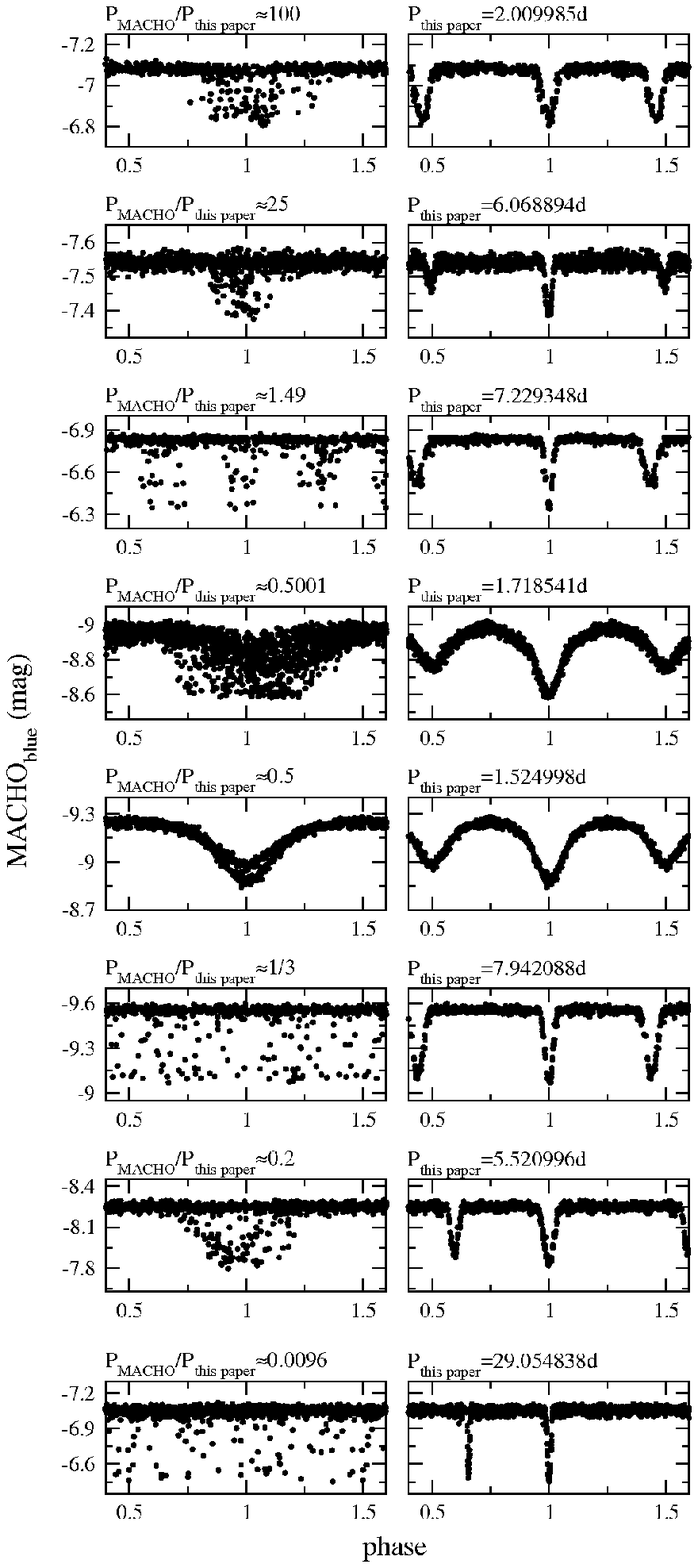}
\caption[]{\label{examples} Examples of typical period ratios. In each case,
the left panel shows the data phased with the MACHO period, while the right
panel shows the final phase diagram with our period.}
\end{figure}

The histogram of the period ratios (Fig.\ \ref{hiscom}) shows that roughly 16\%
of the periods agree. The true period turned out to be the double of the given
MACHO period in about 78\% of the binary sample, while the remaining 6\% have
other ratios. In Fig.\ \ref{examples} we show a few typical examples, plotting
phase diagrams with the MACHO and our periods.

With the corrected phase diagrams, it became possible to re-classify the
sample. As discussed above, pulsating and non-periodic variables were easily
excluded, while eclipsing binaries were visually pre-classified as Algol (EA),
$\beta$ Lyrae (EB) and W UMa (EW) type stars. However, this kind of
classification contains some subjectivity, so we decided to use a more
objective method for this purpose. \citet{ruc93} showed that light curves of W
UMa systems can be quantitatively described using only two coefficients, ${\rm
a_{2}}$ and ${\rm a_{4}}$, of the cosine decomposition ${\rm \sum a_{i} \cos
(2\pi i \varphi)}$. \citet{poj02} tested the behavior of semidetached and
detached configurations in the ${\rm a_{2} - a_{4}}$ plane by decomposing
theoretical light curves into Fourier coefficients. He found that in most cases
contact, semi-detached and detached configurations can be distinguished quite
accurately. We chose this method because it can be easily implemented for a
large set of light curves.

\begin{figure}  
\epsscale{1.15}
\plotone{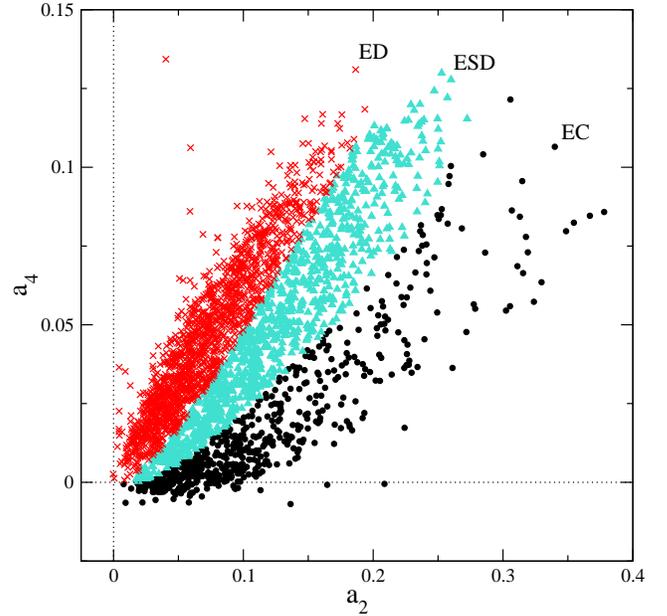}   
\caption[]{\label{fou}Classification of LMC eclipsing binaries in the Fourier
coefficients plane ${\rm a_{2} - a_{4}}$.} 
\end{figure}

Following the definitions by \citet{poj02}, we decomposed every light curve in
the following form:

\begin{equation}
m(\varphi)=m_{0}+ \sum_{i=1}^{4} \left( a_{i} \cos \left( 2\pi i \varphi \right) +
b_{i} \sin \left( 2\pi i  \varphi \right) \right)
\end{equation}

\noindent where $m(\varphi)$ is the phased light curve, $m_{0}$ is the mean
magnitude, while the zero point of the phase corresponds to the primary
minimum. The resulting distribution in the ${\rm a_{2} - a_{4}}$ plane, based
on the MACHO blue data, is shown in Fig.\ \ref{fou}. Using the boundary lines
of \citet{poj02}, we marked the contact (EC), semi-detached (ESD) and detached
(ED) configurations with different symbols.

The sample is dominated by bright main-sequence detached (1482 stars) and
semidetached (937 stars) binaries, while a small fraction consist of contact
systems (612 stars). Many of the latter are foreground Milky Way objects, as
can be shown from their positions in the color-magnitude diagram (Section 3.2).
We list the EC/ESD/ED classification for each star in Table\ \ref{data}.

\begin{table*}
\begin{center}
\caption{\label{data} Representative lines from the full Table 1 (available 
electronically). The (*) symbol in the Type column denotes foreground objects.}
\small
\begin{tabular}{lcccrrcll}
\hline
\hline
 ID number & RA          & Dec.          & V     & V--R  & Period & Epoch      &
 Type & Note\\
            & (J2000.0)   & (J2000.0)     & (mag) & (mag) & (days) & (2400000+) &
	      &  \\
\hline
\noalign{\smallskip}
47.1402.116 & 04:49:21.544 & $-$68:12:31.70 & 18.644 & $-$0.034 & 1.761188 &
50249.7483 & ESD& \\
47.1527.178 & 04:49:28.952 & $-$67:55:24.15 & 19.223 & 0.725 & 12.556414 &
50244.6082 & EC &\\
47.1528.41 & 04:49:34.577 & $-$67:49:48.88 & 17.209 & $-$0.179 & 5.024733 &
50249.8034 & ED& \\
47.1529.60 & 04:49:44.988 & $-$67:48:57.28 & 18.005 & 0.574 & 112.073172 &
50252.7272 & ED& \\
47.1531.32 & 04:49:51.948 & $-$67:41:38.13 & 16.842 & $-$0.159 & 0.459804 &
50249.9699 & EC& \\
47.1530.30 & 04:49:52.060 & $-$67:42:09.84 & 15.519 & $-$0.249 & 1.633056 &
50249.3579 & ED& \\
47.1649.69 & 04:50:16.357 & $-$67:53:06.87 & 17.396 & $-$0.153 & 1.453116 &
50250.0195 & ESD& \\
47.1645.115 & 04:50:21.514 & $-$68:07:37.01 & 18.048 & $-$0.163 & 1.711124 &
50249.3182 & ED& \\
47.1647.143 & 04:50:33.106 & $-$67:58:44.94 & 18.440 & $-$0.186 & 1.620856 &
50250.1698 & ED& \\
47.1647.285 & 04:50:34.347 & $-$67:59:24.49 & 19.653 & 0.082 & 1.387449 &
50249.4512 & ESD& \\
11.9720.84 & 05:40:08.972 & $-$70:15:03.29 & 17.506 & 0.953 & 4.607095 &
50248.2811 & EC(*)& \\
\noalign{\smallskip}
\hline
\end{tabular}
\end{center}
\end{table*}

\section{General properties}

After the visual inspection, we identified 3031 stars as eclipsing binaries,
which is about 44\% of the downloaded sample. We emphasize that our sample
probably does not contain all LMC eclipsing binaries observed by the MACHO
project. We assume that there might be eclipsing binaries classified as other
variable types (such as RR Lyraes, Cepheids, Semiregulars, etc.). For example,
\citet{fac05} studied orbital circularization of LMC eclipsing binaries
presenting a new sample of 4576 stars; however, that sample is not available to
us. With this caveat, we discuss the main properties of the sample.

\subsection{Period distribution}

The most fundamental parameter of binary stars is the orbital period, whose
distribution can be of great aid in understanding formation and evolution of
close binaries. Eclipse detection is highly influenced by the orbital
period, because the wider the separation of components, the longer the orbital
period gets, implying that eclipses will be seen in narrower range of the
inclination angle. Hence the chances to detect eclipses are much higher for
short orbital periods. Well before the microlensing projects, \citet{far78}
and \citet{giu83} found a multi-modal period distribution in the Milky Way.
Recent studies of eclipsing binaries in the Small and the Large Magellanic
Clouds revealed an overall similarity in the period distributions, peaking
between 1 and 2 days \citep{alc97,wyr03,wyr04,dev05}.

\begin{figure}
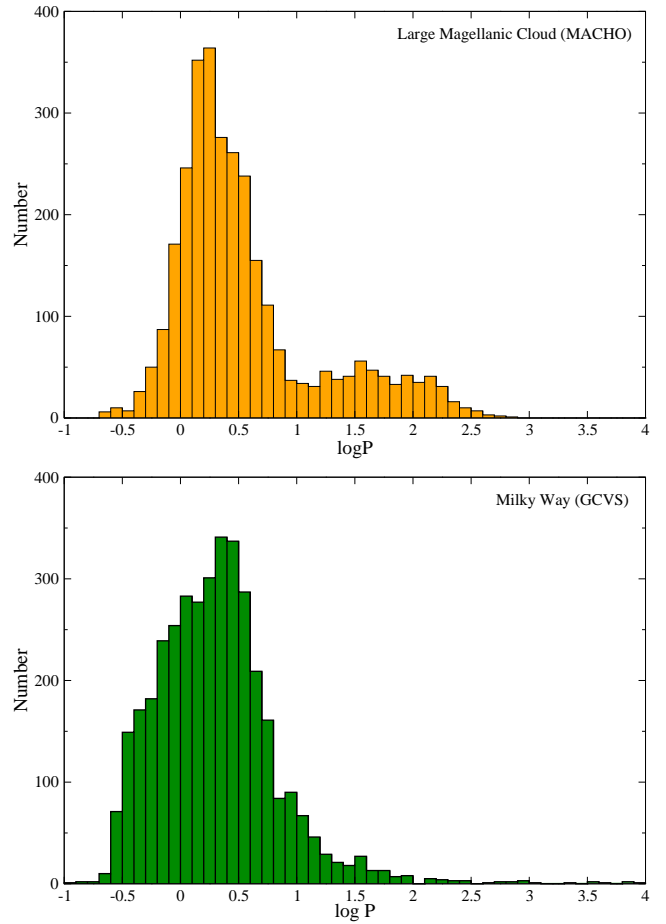

\includegraphics[width=8.5cm]{macho.eps}   
\vskip1mm
\includegraphics[width=8.5cm]{gcvs.eps}
\caption[]{\label{period} Upper panel: period distribution of eclipsing
binaries in the MACHO  database. Lower panel: histogram of orbital periods of
eclipsing binaries in the General Catalog of Variable Stars.}
\end{figure}

For a direct comparison with the OGLE sample for the LMC and the SMC, we
downloaded periods of eclipsing binaries from the OGLE internet
archive\footnote {http://bulge.astro.princeton.edu/~ogle}. Our period
distribution (Fig.\ \ref{period}) is in a good agreement with that of the OGLE
sample of LMC and SMC stars \citep{wyr03,wyr04}: the majority of systems have
short periods, peaking between 1 and 2 days, and roughly 20\% of stars have
periods longer than 10 days, which is consistent for both the SMC and the LMC. 

The distribution shown in the upper panel of Fig.\ \ref{period} is also similar
to that of \citet{dev05}, which was based on the OGLE observations of the
Galactic Bulge. However, we do not confirm his conclusion that the peak at
${\rm \sim 100}$ days is due to pulsating red giant stars. While \citet{dev05}
used single filtered I-band data of the OGLE project, thus having no
information on the color variations, the two-color MACHO observations clearly
showed that those long-period variables have only slight color changes, which
can be explained by the non-uniform temperature distribution on the surface of
the non-spherical components.

The lower panel of Fig.\ \ref{period} shows the histogram of all eclipsing
binaries in the General Catalog of Variable Stars (\citet{kho8588} with all the
recent updates available for download from the GCVS
website\footnote{http://www.sai.msu.su/groups/cluster/gcvs/gcvs/}). Compared to
the upper panel of Fig.\ \ref{period}, the distributions are different in two
period ranges. For shorter periods, there is an excess of stars in the GCVS,
which are all short-period main-sequence binaries that have fainter absolute
magnitudes than the MACHO limit in the LMC. On the other hand, for periods
longer than 40 days, there is a lack of stars in the GCVS, which we interpret
as caused by a selection effect: these systems need years of observations
before classification and period determination, which was hardly possible
before the development of automatic all-sky survey projects like the All-Sky
Automated Survey \citep[ASAS,][]{poj02}. It also implies that there are
considerable number of bright long-period eclipsing binaries waiting for
discovery.

\begin{figure}
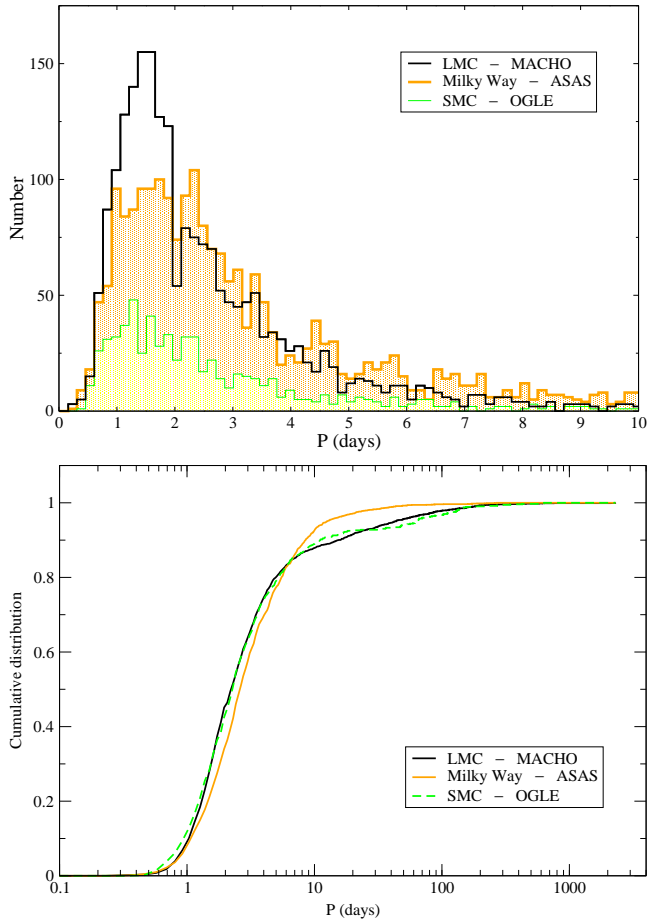
   
\includegraphics[width=8.5cm]{moa.eps}   
\vskip1mm
\includegraphics[width=8.5cm]{cumdist.eps}
\caption[]{\label{moa} Upper panel: Orbital period distributions of detached
and semi-detached eclipsing binaries in the Milky Way and the Magellanic
Clouds. Lower panel: The normalized cummulative distributions.}  
\end{figure}

To make further comparison, we took detached and semi-detached binaries in the
Galaxy observed by the ASAS project and compared their period distribution with
detached and semi-detached binaries in the SMC and LMC. In the upper panel of
Fig.\ \ref{moa} we plotted the period distribution of orbital periods shorter
than 10 days, which shows an overall similarity for the three galaxies. This
diagram also shows the presence of a selection effect in the data: there is a
noticeable dip at P=2d in the histograms of the MACHO and the ASAS data, which
reveals that stars with exactly 1-d periods (i.e. the half of the true ones)
were discarded during the initial analyses. For example, based on the shape of
the histogram, about 50 MACHO eclipsing binaries with ${\rm P\sim2d}$ are
missing from the sample, presumably due to deliberate exclusion from the data.

In the lower panel of Fig.\ \ref{moa} we plot the cumulative period
distribution for the three samples. While the LMC and SMC samples have
virtually indistinguishable distributions, the Milky Way shows an excess of
stars around 10 days. A two-sample Kolmogorov-Smirnov test showed that the
difference between the LMC and the SMC samples is insignificant, which confirms
that the MACHO and OGLE selections of detached and semi-detached binaries were
similarly complete. On the other hand the probability that the ASAS stars have
the same distribution is only $10^{-11}\%$. This is a simple consequence of the
different sampling: while MACHO and OGLE data represent magnitude-limited
samples in absolute magnitudes, covering the brightest $\sim6$ mag of the CMD,
the ASAS database contains a broad mixture of stars in the Milky Way with a
much wider range in absolute magnitude.

It is also apparent in Fig.\ \ref{moa} that, although the long-period tails of
the ASAS and MACHO histograms overlap very well, the short-period peak in the
LMC data is higher because of the different distributions of stars in the
samples. Whereas the MACHO data cover the upper part of the Hertzsprung-Russell
diagram with a relatively bright limit in absolute magnitude, the ASAS sample
contains a broader mixture of stars in the galactic neighborhood, so that
short-period B-type systems have a smaller relative contribution.

\subsection{The Color-Magnitude Diagram and Period-Luminosity Relations}

\begin{figure}  
\epsscale{1.15}
\plotone{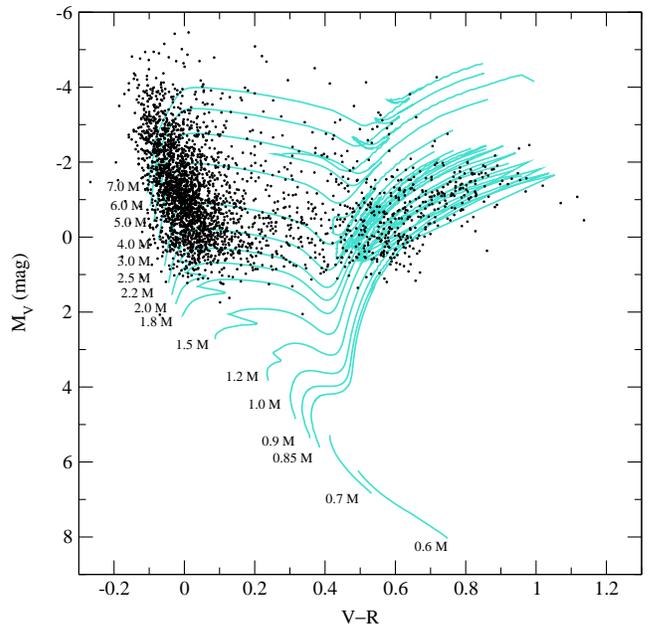}
\caption[]{\label{cmd}The color-magnitude diagram of LMC eclipsing binaries
with metal-poor evolutionary tracks of single stars overlaid (models taken from
\citet{cas03}).} 
\end{figure}

In order to construct the Color-Magnitude Diagram (CMD), we converted the
observed MACHO blue and red magnitudes into  Kron-Cousins V and R using the
equations derived by \citet{alc99}. The resulting diagram, after translating
apparent magnitudes outside the eclipses to absolute magnitudes using
${\rm \mu(LMC)=18.50}$, is shown in Fig.\ \ref{cmd}. We also show stellar
evolutionary tracks by \citet{cas03}. Most of the stars are moderately massive
near-main sequence stars, while the red giant branch of evolved stars is also
clearly recognisable. The latter is dominated by first-ascent red giants and a
few Asymptotic Giant Branch stars \citep{alc00}. 

\begin{figure*}
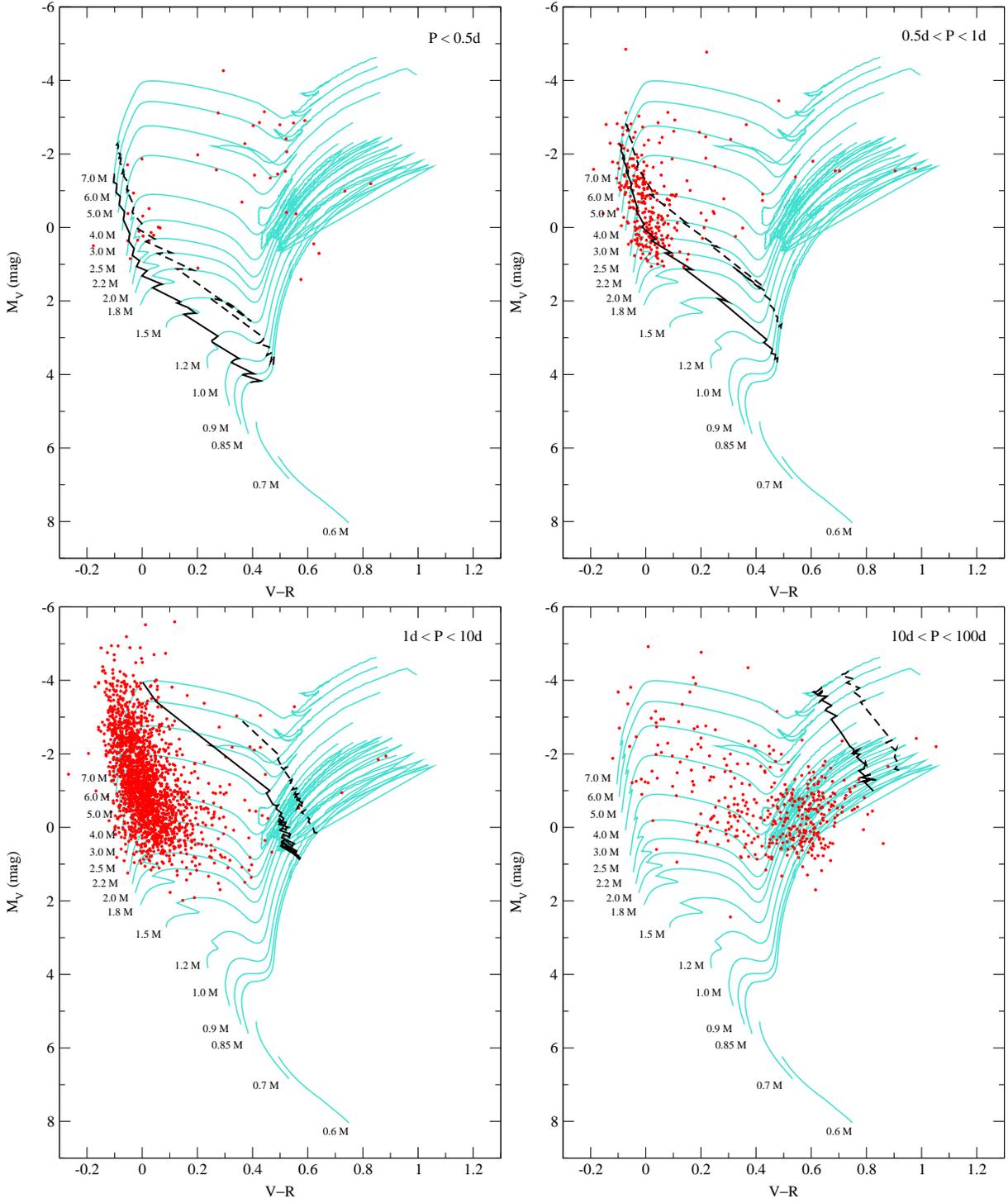

\begin{center}
\includegraphics[width=8cm]{05track.eps}   
\includegraphics[width=8cm]{1track.eps}   

\includegraphics[width=8cm]{10track.eps}   
\includegraphics[width=8cm]{100track.eps}   
\end{center}
\caption[]{\label{tracks} Eclipsing binaries in four period ranges and estimated
locations of the minimum orbital periods. Stars upper and right of the dashed
lines have positions that are not compatible with the LMC membership.}
\end{figure*}

In Fig.\ \ref{tracks} we plot the CMD for four orbital period ranges. These
plots were used for cleaning the sample of the foreground objects, adopting the
following simple considerations. First, the spread of distances of eclipsing
binaries in the LMC is negligible compared to the distance to the LMC. This
means that we can determine the absolute magnitude for any given object within
$\pm$0.1 mag \citep[e.g.][]{nic04,lah05}.  Second, we assumed that the mean
position of a binary star in the CMD can be approximated by the location of the
brighter component. In the case of two identical components, this assumption
means a 0.75 mag fainter absolute magnitude, however, as we proceeded, this was
still a useful simplification in estimating foreground contamination.

Theoretical evolutionary tracks give the basic physical parameters of a single
star, as well as color and magnitude information. From the physical parameters
we can calculate the minimum orbital period for any given mass, temperature and
luminosity, applying simple relations. Combining the Stefan-Boltzmann law:
$L/L_{\odot}=\left(R/R_{\odot}\right)^{2} \left(T/T_{\odot}\right)^{4}$ and
Kepler's third law: $a^{3}/P^{2}=G\left(M_{1}+M_{2}\right)/4\cdot\pi^{2}$,
we can calculate the minimum orbital period of a system for two extreme cases: 

{\it Case 1.}: two identical components, for which the minimum orbital period
occurs when $a=2R$ and $M_{1}+M_{2}=2M$; the minimal orbital period is

\begin{equation} 
P_{\rm min}(1)=4\pi\sqrt{\frac{R^{3}}{GM}}
\label{minper1}
\end{equation}

{\it Case 2.}: a negligible secondary component, for which the minimum orbital
period occurs when $a=R$ and $M_{1}+M_{2}=M$; the minimum orbital period is

\begin{equation} 
P_{\rm min}(1)=2\pi\sqrt{\frac{R^{3}}{GM}}
\label{minper2}
\end{equation}

\noindent ($M$ is the mass of a model point of the isochrone, while $R$ is
calculated from $L$ and $T$).

Using Eq.~(\ref{minper1}) and Eq.~(\ref{minper2}) and theoretical stellar
evolutionary tracks, we can determine contours in the CMD, on which the minimum
orbital period has a given value. This is shown in the four panels of Fig.\
\ref{tracks} by the thick solid lines (Case 1) and dashed lines (Case 2). As
representative limits, we determined the locations where the minimum orbital
period is 0.5, 1, 10 and 100 days.

We have selected the foreground objects as follows. As Fig.\ \ref{tracks}
shows, moving up and right in the CMD, the minimum orbital period gets longer.
Therefore, if a star with $P_{\rm orb} \leq 0.5{\rm d}$ is located above the
limiting line of $P_{\rm min} = 0.5{\rm d}$, it must be in the foreground of
the LMC. This is shown in the upper left panel of Fig.\ \ref{tracks}. As
expected, most of these short-period reddish eclipsing binaries are Galactic W
UMa-type stars, many magnitudes above the calculated period lines (which means
the adopted simplifications do not affect the conclusions).

For longer periods, only a few stars are clearly foreground objects (most
notably in the upper right panel of Fig.\ \ref{tracks}), while for the
majority, positions in the CMD are compatible with the LMC membership. We
flagged all obvious foreground stars, being those located at least 1 mag above
the dashed lines, in the last column of Table\ \ref{data} (54 stars in total).

\begin{figure}  
\epsscale{1.15}
\plotone{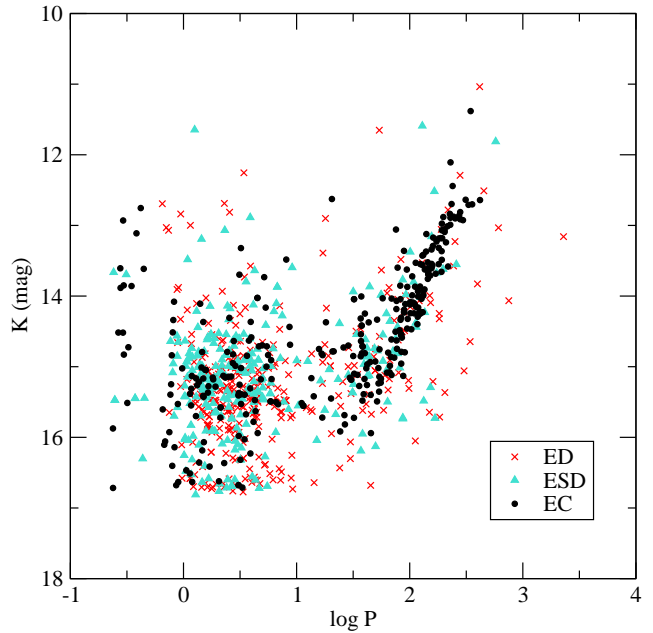}   
\caption[]{\label{logp}Period-K magnitude relation for eclipsing binaries in 
the LMC.} 
\end{figure}

Finally, we briefly examine the period-luminosity (PL) relation of stars in the
sample, for which we found near-infrared K-magnitudes in the 2MASS Point-Source
Catalog. In \citet{der06} we already discussed how the red giant eclipsing
binaries/ellipsoidal variables can be fitted with a simple model using
Roche-lobe geometry. Here we examine the correlation between the period and the
K-magnitude for the detached and semi-detached binaries. This is shown in Fig.\
\ref{logp}, where the ED/ESD/EC classes are plotted with different symbols. As
expected, detached binaries are spreaded uniformly, while longer period
semi-detached systems may follow a loose correlation (the correlation
coefficient for ${\rm P>10d}$ is $\sim 0.58$), but it is not as tight as for
long-period contact binaries (whose correlation coefficient is $\sim 0.86$).
Systems further away from the main correlation line might be foreground
stars left unidentified as such in the CMD analysis.

\section{Summary}

In this paper we present the analysis of online light curves of eclipsing
binaries in the Large Magellanic Cloud monitored by the MACHO project. We
downloaded the data of all 6833 stars classified as eclipsing binaries from the
Variable Star Catalog Retrieval form of the project.

We re-determined the periods for every star and reclassified all stars based on
their light curve shape. As a result, 3031 stars remained as eclipsing binary,
while the rest of the sample were RR Lyrae stars, Cepheids or long-period
pulsators. For the binary sample, we showed that roughly 16\%  of the periods
agreed with those given in the catalog. For almost three-quarter of the sample,
the catalog periods turned out to be the half of the real ones. 

We calculated the period histogram, which shows bimodal feature on logarithmic
scale. The maximum of the distribution is at binaries with periods between 1
and 2 days, while roughly 20\% of the sample (about 600 stars) have period
between 10 and 200 days. We compared our period distribution for detached
eclipsing binaries, with those of the SMC and the Milky Way and we found a good
agreement in the most common periods in the studied galaxies.

For a more objective classification, we used cosine decomposition of the light
curves into Fourier coefficients, where contact, semi-detached and detached
configurations can be distinguished in ${\rm a_{2} - a_{4}}$ plane. The sample
is dominated by bright main-sequence detached and semi-detached systems, while
a small fraction consists of contact binaries. However, many of the latter are
foreground Milky Way objects, which can be shown from their position in the
color-magnitude diagram. Most of the stars with periods shorter than 0.5 days
are galactic W UMa systems. For longer period systems, we found a few obviously
foreground objects and they are flagged in the last column of Table\
\ref{data}. 

The presented database opens up a new avenue into using the MACHO database for
studying secular variations over a time-scale of a decade. This includes a
search for systems with apsidal motion (recognizable through phase shifts of
the primary and secondary minima in the opposite direction), tertiary
components (indicated by cyclic phase shifts of the primary and secondary
minima in the same direction) and eclipsing binaries with pulsating components
(revealed by periodic secondary brightness fluctuations). The latter would be
particularly interesting if evidence for tidally induced oscillations could be
found in a larger sample of stars. With the OGLE-III project still taking data,
after identifying these theoretically important test objects, it would become
possible to extend the full time-coverage of observations up to 14-15 years. In
a companion paper we will discuss the measured period changes for the whole
sample of 3031 eclipsing binaries. To help the interested researchers, we have
collected the original MACHO blue and red light curve data in a single
compressed data file that can be accessed as an electronic appendix to this
paper.

\acknowledgments

We are greatful to the anonymous referee for his/her comments that improved the
paper.

AD is supported by an Australian Postgraduate Award of the Australian
Department of Education, Science and Training. LLK is supported by a University
of Sydney Postdoctoral Research Fellowship.  This work has been supported by
the Australian Research Council. 

The NASA ADS Abstract Service was used to access data and references. 

This paper utilizes public domain data obtained by the MACHO Project, jointly
funded by the US Department of Energy through the University of California,
Lawrence Livermore National Laboratory under contract No. W-7405-Eng-48, by the
National Science Foundation through the Center for Particle Astrophysics of the
University of California under cooperative agreement AST-8809616, and by the
Mount Stromlo and Siding Spring Observatory, part of the Australian National
University.


\begin{thebibliography}{}
\expandafter\ifx\csname natexlab\endcsname\relax\def\natexlab#1{#1}\fi
\expandafter\ifx\csname url\endcsname\relax
  \def\url#1{{\tt #1}}\fi


\bibitem[Alcock et al.(1993)]{alc93}
  Alcock, C., et al., 1993, ASPC, 43, 291

\bibitem[Alcock et al.(1997)]{alc97}
  Alcock, C., et al., 1997, \aj, 114, 326
  
\bibitem[Alcock et al.(1999)]{alc99}
  Alcock, C., et al., 1999, \pasp, 111, 1539
  
\bibitem[Alcock et al.(2000)]{alc00}
  Alcock, C., et al., 2000, \aj, 119, 2194

\bibitem[Aubourg et al.(1995)]{aub95}
  Aubourg, E., et al., 1995, \aap, 301, 1

\bibitem[Castellani et al.(2003)]{cas03}
  Castellani, V., Degl'Innocenti, S., Marconi, M., Prada Moroni, P. G., \& 
  Sestito, P., 2003, \aap, 404, 645

\bibitem[Cook(1995)]{coo95}
  Cook, K. H., Alcock, C., Allsman, H. A., et al. 1995, in Astrophysical
  Applications of Stellar Pulsation, ed. R. S. Stobie, \& P. A. Whitelock, ASP 
  Conf. Ser., 83, 221 
  
\bibitem[Clarke(2002)]{cla02}
  Clarke, D., 2002, \aap, 386, 763

\bibitem[Derekas et al.(2006)]{der06} 
  Derekas, A.,  Kiss, L. L., Bedding, T. R., Kjeldsen, H., Lah, P., \& Szab\'o,
  Gy. M., 2006, \apj, 650, L55

\bibitem[Devor(2005)]{dev05}
  Devor, J., 2005, \apj, 628, 411

\bibitem[Faccioli et al.(2005)]{fac05}
  Faccioli, L., Alcock, C., Cook, K., Prochter, G., \& Syphers, D., 2005, in 
  Tidal evolution and oscillations in binary stars, ASP Ser., ed. A. Claret, A. 
  Giménez, \& J.-P. Zahn, 333, 75

\bibitem[Farinella \& Paolicchi(1978)]{far78}
  Farinella, P., \& Paolicchi, P., 1978, \apss, 54, 389

\bibitem[Fitzpatrick et al.(2002)]{fit02}
  Fitzpatrick, E. L., Ribas, I., Guinan, E. F., DeWarf, L. E., Maloney, F. P.,
  \& Massa, D., 2002, \apj, 564, 260

\bibitem[Giuricin et al.(1983)]{giu83}
  Giuricin, G., Mardirossian, F., \& Mezzetti, M., 1983, \aap, 119, 218

\bibitem[Guinan et al.(1998)]{gui98}
  Guinan, E. F., et al., 1998, \apj, 509, 21
  
\bibitem[Johnson et al.(2004)]{joh04}
  Johnson, A., Whelan, D., Edinger, B., Bailey, B., Smith, K., Malmrose, M., \&
  Palen, S. E., 2004, BAAS, 36, 740

\bibitem[Kholopov et al.(1985-1988)]{kho8588}
  Kholopov et al. 1985-1988, General Catalogue of Variable Stars, Vol. III,
  Nauka, Moscow

\bibitem[Lah et al.(2005)]{lah05}
  Lah, P., Kiss, L. L., \& Bedding, T. R., 2005, \mnras, 359, 42L

\bibitem[Lafler \& Kinman(1965)]{laf65}
  Lafler, J., \& Kinman, T. D., 1965, \apjs, 11, 216

\bibitem[Maceroni \& Rucinski(1999)]{mac99}
  Maceroni, C., \& Rucinksi, S. M., 1999, \aj, 118, 1819

\bibitem[Mazeh et al.(2006)]{maz06}
  Mazeh, T., Tamuz, O., North, P., 2006, \mnras, 367, 1531

\bibitem[Michalska \& Pigulski(2005)]{mic05}
  Michalska, G., \& Pigulski, A., 2005, \aap, 434, 89

\bibitem[Nikolaev et al.(2004)]{nic04}
  Nikolaev, S., Drake, A. J., Keller, S. C., Cook, K. H., Dalal, N., Griest, K.,
  Welch, D. L., \& Kanbur, S. M., 2004, \apj, 601, 260

\bibitem[Paczy\'nski et al.(2006)]{pac06}
  Paczy\'nski, B., Szczygiel, D., Pilecki, B., \& Pojma\'nski, G., 2006, \mnras,
  368, 1311

\bibitem[Palen \& Armstrong(2003)]{pal03}
  Palen, S., \& Armstrong, J. C., 2003, BAAS, 35, 1222

\bibitem[Pojma\'nski(2002)]{poj02}
  Pojma\'nski, G., 2002, \actaa, 52, 397
  
\bibitem[Ribas et al.(2000)]{rib00}
  Ribas, I., et al., 2000, \apj, 528, 692

\bibitem[Ribas et al.(2002)]{rib02}
  Ribas, I., Fitzpatrick, E. L., Maloney, F. P., Guinan, E. F., Udalski, A., 
  2002, \apj, 574, 771

\bibitem[Rucinski(1993)]{ruc93}
  Rucinski, S. M., 1993, \pasp, 105, 1433

\bibitem[Rucinski(1997a)]{ruc97a}
  Rucinski, S. M., 1997a, \aj, 113, 407

\bibitem[Rucinski(1997b)]{ruc97b}
  Rucinski, S. M., 1997b, \aj, 113, 1112

\bibitem[Rucinski(1998)]{ruc98}
  Rucinski, S. M., 1998, \aj, 115, 1135


\bibitem[Rucinski \& Maceroni (2001)]{ruc01}
  Rucinski, S. M., \& Maceroni, C., 2001, \aj, 121, 254

\bibitem[Salaris \& Groenewegen(2002)]{sal02}
  Salaris, M., \& Groenewegen, M. A. T., 2002, \aap, 381, 440

\bibitem[Stellingwerf(1978)]{ste78}
  Stellingwerf, R. F., 1978, \apj, 224, 953

\bibitem[Tamuz et al.(2006)]{tam06}
  Tamuz, O., Mazeh, T., North, P., 2006, \mnras, 367, 1521

\bibitem[Udalski et al.(1997)]{uda97}
  Udalski, A., Kubiak, M., \& Szyma{\'n}ski, M., 1997, \actaa, 47, 319

\bibitem[Wyithe \& Wilson(2001)]{wyi01}
  Wyithe, J. S. B., \& Wilson, R. E., 2001, \apj, 559, 260

\bibitem[Wyrzykowski et al.(2003)]{wyr03}
  Wyrzykowski, L., et al., 2003, \actaa, 53, 1

\bibitem[Wyrzykowski et al.(2004)]{wyr04}
  Wyrzykowski, L., et al., 2004, \actaa, 54, 1

\end{thebibliography}
\end{document}